\documentclass[aps,showpacs,preprintnumbers,amsmath,amssymb]{revtex4}

\UseRawInputEncoding
\usepackage{graphicx}
\usepackage{epstopdf}
\usepackage{color}
\usepackage{subfig}
\usepackage{aasmacros}
\usepackage{amsmath,amssymb,amsfonts}
\usepackage{accents}

\begin{document}
    \captionsetup{justification=raggedright,singlelinecheck=false}

    \baselineskip=0.8cm
    \title{\bf Bright ring features and polarization structures in  Kerr-Sen black hole images of Sagittarius A* illuminated by radiatively inefficient accretion flows}

    \author{Hao Yin$^{1}$,
    Songbai Chen$^{1,2}$\footnote{Corresponding author: csb3752@hunnu.edu.cn},
    Jiliang Jing$^{1,2}$ \footnote{jljing@hunnu.edu.cn}}
    \affiliation{$^1$Department of Physics, Institute of Interdisciplinary Studies, Hunan Research Center of the Basic Discipline for Quantum Effects and Quantum Technologies, Key Laboratory of Low Dimensional Quantum Structures
    and Quantum Control of Ministry of Education, Synergetic Innovation Center for Quantum Effects and Applications, Hunan
    Normal University,  Changsha, Hunan 410081, People's Republic of China
    \\
    $ ^2$Center for Gravitation and Cosmology, College of Physical Science and Technology, Yangzhou University, Yangzhou 225009, People's Republic of China}

    \begin{abstract}
    \baselineskip=0.6 cm
    \begin{center}
    {\bf Abstract}
    \end{center}
    Using general relativistic radiative transfer (GRRT) simulations, we investigate the bright ring features and polarization structures in images of the Kerr-Sen black hole associated with Sgr A*, as illuminated by 230 GHz thermal synchrotron emission from radiatively inefficient accretion flows (RIAF). Our findings reveal that an increase in the dilaton parameter leads to a shrinking of the bright ring, accompanied by enhancements in both its width and brightness. As the disk thickness grows, the bright ring's diameter and width both decrease. The brightness enhancement induced by the disk thickness is less prominent than that driven by the dilaton parameter. Comparing with  the Event Horizon Telescope (EHT) observational data of SgrA*, we present the allowed ranges of black hole parameters, and find that effects of the disk thickness on the allowed parameter space are stronger than those of the observer's inclination. Furthermore, we analyze the coefficient $\beta_2$, associated with the two-fold rotational symmetry of the electric vector position angles (EVPA),
    to probe the polarization structure of the black hole images, and reveal that effects of the disk thickness on $\beta_2$ are much weaker than those from the dilaton parameter.

    \end{abstract}

    \pacs{ 04.70.Dy, 95.30.Sf, 97.60.Lf }
    \maketitle
    \newpage

    \section{Introduction}

    The horizon-scale images of M87* and Sgr A* captured by the Event Horizon Telescope (EHT) have opened an unprecedented window for testing gravity in strong-field regimes \cite{sgraI,EHTsgraVI,m87I,EventHorizonTelescope:2019ths}. The brightness distributions and polarization patterns in black hole images contain a wealth of information about electromagnetic emissions in the vicinity of black holes, offering a powerful means to probe the matter distribution and accretion processes surrounding these celestial objects. This is invaluable for gaining insights into the physics in the strong-field regions and verifying theories of gravity \cite{Psaltis_2020,Vagnozzi_2023}. Consequently, extensive efforts have been dedicated to investigating the images of various black hole systems \cite{Qin:2023nog,zhang2024imageskerrmogblackholes,Chen:2022scf,Wang:2023jop}.

    The bright ring of emission is one of the most important ingredients in the images captured by EHT, which arises mainly from the strong gravitational lensing of synchrotron emission produced by the accretion flow surrounding the black hole \cite{Desire_2025}. Actually, this bright ring can be decomposed into finer substructures \cite{Johnson:2019ljv}, most notably the photon ring which originates from photons completing an infinite number of half-orbits around the black hole. Theoretically, the position of the light ring is entirely determined by the spacetime properties of the black hole and is independent of the properties of the accretion disk \cite{Johnson_2020}. However, it is difficult to be directly observed in the black hole images because the photon ring is almost overlapped with other subrings and its  brightness is obscured by other lower-order images. In addition, the current observational precision also does not allow us to detect the light ring in real astronomical observations and the most likely thing to be observed is the bright ring. Because the bright ring depends not only on black hole parameters but also on the properties of the accretion disk around the black hole, it is necessary to reliably extract the bright ring from the blurred black hole images with finite resolution. This is because the size and brightness of the bright ring can facilitate the capture of key characteristics of the black hole and its accretion disk. With the REx algorithm \cite{chael2019simulating}, the EHT Collaboration found that the bright ring diameter in the Kerr black hole case remains robust across different imaging methods, while several other features of the ring, such as its width, orientation, asymmetry, and fractional central brightness, depend on the imaging pipeline \cite{2019ApJ...875L...4E, sgraIV,EventHorizonTelescope:2019ggy,achour2025blackholephotonring}.

    Polarization patterns are another important ingredients in black hole images observed by EHT because
    because they carry unique information about the magnetic fields and plasma dynamics in the extreme environment surrounding the black hole. The structure of near-horizon linear polarization in EHT images can be quantified by the second azimuthal Fourier mode $\beta_2$ of the linear polarization image \cite{ehtVII,m87VIII,Palumbo_2020}, while the corresponding helicity of the spiral of polarization vectors can be characterized by its phase $\angle\beta_2$. The value of $\angle\beta_2$  can also be applied to constrain the direction of the near-horizon electromagnetic energy flow of the black hole M87* \cite{ehtVII}. In addition, $\beta_2$ has been shown to not only distinguish whether  the magnetic field around the black hole is in the magnetically arrested disk (MAD) or standard and normal evolution (SANE) states, but also further discriminate between different spin values within MAD models~\cite{Palumbo_2020}. Moreover, it has been demonstrated that $\angle \beta_2$ shows some sensitivity to the spacetime geometry~\cite{jiang2023shadowsloopquantumblack}. For the black hole Sgr A*, the inclination angle is estimated to be around $30^\circ$ \cite{2024ApJ...964L..26E,EHTsgraV,sgraI}. This low inclination is sufficient to yield rotational symmetry that is accessible to $\beta_2$ analysis.

    The bright ring features and polarization structures in black hole images also depend on the magnetized accretion flows around the black hole. In general, the full dynamics of accretion flows must resort to general relativistic magnetohydrodynamics  (GRMHD) simulations with high precision \cite{2022ApJ...930L..16E,2019m87V}. However, their high computational cost renders a systematic exploration of the vast parameter space facing enormous challenges. A semi-analytic radiatively inefficient accretion flow (RIAF) model is a feasible alternative for describing magnetized accretion flows near black holes \cite{Yuan_2003,Pu_2018}. For the current EHT targets, M87* and Sgr A*, the
   surrounding hot plasma is considered to be part of a radiatively inefficient accretion flow, which is typically hot, geometrically thick, and optically thin at an observing frequency of 230 GHz. The corresponding horizon-scale images have been analyzed for Kerr black hole \cite{jiang2023shadowsloopquantumblack,Yan:2025mlg,Pu_2018,Pu:2016qak}.  Therefore, it is highly valuable to investigate the impact of black hole parameters and accretion disk thickness on the bright ring features and polarization structures for other black holes within the RIAF framework.

    In this paper, we focus on the Kerr-Sen black hole and study effects of dilaton parameter together with accretion disk thickness on bright ring features and polarization structures in the black hole images. The  motivation is based on the following aspects: Firstly, the Kerr-Sen black hole is a solution in string theory, which is widely regarded as a promising candidate for the unified theory of everything. Secondly, the Kerr-Sen metric exhibits substantial differences in its intrinsic geometric properties from the Kerr-Newman solution in general relativity, despite their strong resemblance. Furthermore, investigating this background's impact on black hole images  not only provides a possibility for testing string theory but also offers a scope to verify the Kerr hypothesis, especially given the future development of high-resolution imaging via very long baseline interferometry (VLBI).
    The Kerr-Sen black hole carries charge stemming from the $U(1)$ gauge field in string theory, with this charge accompanied by a dilaton field originating from string compactifications. Unlike the standard electromagnetic charge, the $U(1)$ charge of Kerr-Sen spacetime essentially originates from the axion-photon coupling rather than infalling charged particles. When the charge vanishes, one can find the field corresponding to both the axion and the dilaton disappear, a conclusion that can also be obtained from Eq.(\ref{xidilation}).
    The black holes in astronomical environments are commonly regarded as being electrically neutral, as oppositely charged particles captured from their surroundings are expected to rapidly screen their charge. However, it is shown that a residual charge may still persist in certain accretion flow scenarios \cite{1978PhRvD..17.1518D}. Meanwhile, observations of the supermassive black hole at the center of the Milky Way suggest that Sgr A* \cite{2018MNRAS.480.4408Z,2021PhRvL.126d1103B,2021PhRvD.103j4047K} may indeed possess a small charge, with an upper observational limit $q<3\times 10^{8}C$ \cite{2018MNRAS.480.4408Z}. Thus, it is a worthwhile endeavor to search for signatures of these charges, which are accompanied by dilaton fields,  in available astrophysical observations.
     Studies of the Kerr-Sen black hole have shown that the dilaton parameter does indeed affect the shape of its shadow \cite{Wei:2013kza,Guo_2020,Hioki_2008,Uniyal_2017,Tsukamoto:2024asy,Yagi:2023eap,BenAchour:2025uzp}. However, these studies have not addressed the optical morphology of a Kerr-Sen black hole surrounded by RIAFs. It is therefore important to investigate whether the dilaton parameter influences observable characteristics including the bright ring and polarization pattern in black hole images illuminated by such accretion flows, particularly in the blurred ones.

    This paper is organized as follows: In Sec. II, we briefly introduce the Kerr-Sen black hole and the RIAF model. In Sec. III, we adopt the REx algorithm to extract the features of the bright ring from black hole images generated by GRRT, and then investigate the influences of the dilaton parameter and disk thickness on these images. In Sec. IV, we apply the  $\beta_2$  mode to analyze how the dilaton parameter and disk thickness affect the polarization structures in the images of Kerr-Sen black holes. Finally, we present a summary.

    \section{MODELING RIAF DYNAMICS around Kerr-Sen black holes}
    \label{sec:2}
    The Kerr-Sen black hole  is a stationary and axisymmetric solution in Einstein-Maxell-Dilaton-Axion theory,  whose metric has the form in the Boyer-Lindquist coordinates \cite{Bernard_2016,Ghezelbash_2013,PhysRevLett.74.1276,Sen_1992}
    \begin{eqnarray}
        \label{metric}
        \begin{aligned}
        \mathrm{d}s^{2} & =-\left(1-\frac{2Mr}{\tilde{\Sigma}}\right)\mathrm{d}t^{2}+\frac{\tilde{\Sigma}}{\Delta}\mathrm{d}r^{2}+\tilde{\Sigma}\mathrm{d}\theta^{2}-\frac{4aMr}{\tilde{\Sigma}}\sin^{2}\theta\mathrm{d}t\mathrm{d}\phi \\
         & +\sin^2\theta\mathrm{d}\phi^2\left[r(r+r_2)+a^2+\frac{2Mra^2\sin^2\theta}{\tilde{\Sigma}}\right],
        \end{aligned}
    \end{eqnarray}
    where
    \begin{eqnarray}
    \Delta=r(r+r_2)-2Mr+a^2,\quad\quad\quad \tilde{\Sigma}=r(r+r_2)+a^2\cos^2\theta.
    \end{eqnarray}
   In the Kerr-Sen solution, the axion, dilaton and  $U(1)$ gauge fields admit the explicit forms \cite{Sen_1992}
    \begin{eqnarray}
    \xi = \frac{q^{2}}{M}\frac{a\cos\theta}{r^{2} + a^{2}\cos^{2}\theta},\quad\quad
    e^{2\chi} = \frac{r^{2} + a^{2}\cos^{2}\theta}{r(r + r_{2}) + a^{2}\cos^{2}\theta} ,\quad\quad
     A=\frac{qr}{\tilde{\Sigma}}\bigg(-dt +a \mathrm{sin}^2\theta d\phi\bigg) .\label{xidilation}
    \end{eqnarray}
    Here $M$ and $a$ represent the black hole's mass and angular momentum, respectively. The dilaton parameter $r_2$ is related to the black hole's electric charge $q$ and the inverse string tension $\alpha'$ by $r_2={\frac{\alpha'}{8}}\frac{q^2}{M}$.
    Theoretically, string theory predicts $r_2$ to be of the order of the square of the Planck length $l^2_p$ since the inverse string tension $\alpha' \simeq  l^2_p$ \cite{Polchinski_1998}. Here, we follow the standard phenomenological approach in black hole imaging studies \cite{Mizuno_2018,Roder:2023oqa,Dastan_2016} and treat $r_2$ as a free parameter to investigate its potential  effects deviating from general relativity.  The origin of the electric charge in the Kerr-Sen solution is not from the falling charged particles, but from the axion-photon coupling \cite{Banerjee_2021,Tripathi:2021rwb}.
    As the dilaton parameter $r_2$ vanishes, the metric (\ref{metric}) reduces to the Kerr black hole metric  instead of the Kerr-Newman one, implying that the charge of the Kerr-Sen black hole is essentially different from that of pure electromagnetic origin. When the rotation parameter $a$ disappears, the spacetime is reduced to a static spherically symmetric black hole determined exactly by its mass and electric charge, as well as the asymptotic dilaton field \cite{YAZADJIEV_1999,PhysRevD.43.3140}.
    It is noted that the Kerr-Sen spacetime Eq. (\ref{metric}) can also be derived through a Newman-Janis transformation\cite{Newman:1965tw} on the spherically symmetric solution in pure dilaton-coupled gravity \cite{YAZADJIEV_1999,PhysRevD.43.3140}.
    The radii of black hole horizons satisfy condition $\Delta = 0$ and have the form
    \begin{equation}
        \label{horizon}
        r_{\pm}=M-\frac{r_2}{2}\pm\sqrt{(M-\frac{r_2}{2})^2-a^2},
    \end{equation}
    where $r_+$ and $r_-$, respectively, correspond to the cases of the outer and inner horizons. It is clear that this solution describes a nonextremal black hole for $r_+>r_-$, while a naked singularity appears if $(M-\frac{r_2}{2})^2<a^2$.
    \begin{table}
    \centering
    \begin{tabular}{c c c}
        \hline
        Parameter & Value & Parameter Description\\
        \hline
        $M_{\rm BH}$ & $4.3\times10^6\rm{M}_\odot$ & Black hole mass \\
        $D_{s}$ &  $8.3 \times 10^3 \rm{pc}$ & Distance to the source\\
        $\delta$ & 1.1 & $n_{\rm e}$ power law index\\
        $\gamma$ & 0.84 & $T_{\rm e}$ power law index\\
        $\kappa_{\rm K}$ & $0.5$ & Keplerian parameter\\
        $\kappa_{\rm ff}$ & $0.5$ & Radial infall parameter\\
        $n_{\rm e,0}$ & $\approx10^7\,\rm{cm}^{-3}$ & Number density\\
        $T_{\rm e,0}$ & $\approx10^{11}\,\rm{K}$ & Temperature of electrons\\
        $\nu_{\rm obs}$ & $230$\,GHz & Observing frequency\\
        FOV & $200\,\mu\rm{as}$ & Field of view\\
        \hline
    \end{tabular}
    \caption{Fiducial parameters for Sgr A* and the surrounding RIAFs }
    \label{table:param}
    \end{table}
    Black hole images are jointly determined by both the black hole parameters and the surrounding plasma \cite{Younsi_2023,_zel_2022}.
    In general, the axion, dilaton and electromagnetic fields of Kerr-Sen black hole interact with the disk. However, the high complexity of such interactions makes a comprehensive analysis challenging, even with high-performance computing clusters at one's disposal. For simplicity, in this work, we only consider the influence of the spacetime metric on the accretion disk as in \cite{Mizuno_2018,Fromm:2021flr,Roder:2023oqa}. For the current observed targets of the EHT including M87* and Sgr A*, their surrounding hot plasmas are considered to be part of a RIAF.
    Here, we adopt a semi-analytical optically thin RIAF model \cite{Yuan_2003,Pu_2018} around the Kerr-Sen black hole to perform general relativistic ray-tracing numerical simulations \cite{Noble:2007zx,moscibrodzka2017ipolesemianalyticscheme}.  In this semi-analytical RIAF model, the electron number density and temperature are assumed to follow power-law profiles with radius,
    \begin{align}
    n_{\rm e} &= n_{\rm e,0} \left( r \right)^{-\delta} \exp \left[- \frac{1}{2} (H \tan \theta)^{-2} \right], \label{eq:ne} \\
    T_{\rm e} &= T_{\rm e,0} \left( r \right)^{-\gamma} , \label{eq:Te}
    \end{align}
    where $\delta$ and $\gamma$ are the power-law indices for the electron density and the electron temperature, respectively. The temperature index $\gamma$ is observationally constrained by the spectral energy distribution \cite{Broderick:2010kx} and VLBI brightness temperature measurements \cite{Yfantis:2024eab}. The parameter $H$ describes the geometric thickness of the disk.
    Here, we adopt $\delta = 1.1$ and $\gamma = 0.84$, which are consistent with these observational constraints and previous studies of RIAFs \cite{Pu_2018}. Since the magnetically arrested disk (MAD) model is favored by the current EHT observations \cite{EHTsgraV}, here a MAD-like configuration \cite{Chen_2022,jiang2023shadowsloopquantumblack} is adopted in our simulations.

   To model the four-velocity of the accretion flow, one can interpolate between Keplerian orbital motion and geodesic free fall as described in \cite{Pu_2018,Pu:2016qak}
    \begin{align}
        u^r &= u^r_{\rm K} + \kappa_{\rm ff}(u^r_{\rm ff} - u^r_{\rm K}), \label{eq:ur} \\
        \Omega &= \Omega_{\rm K} + (1-\kappa_{\rm K})(\Omega_{\rm ff} - \Omega_{\rm K}). \label{eq:Omega}
    \end{align}
    where $(u^r, \Omega)$ and $(u^r_{\rm ff}, \Omega_{\rm ff})$ denote the radial component of the four-velocity $u^r$ and the angular velocity $\Omega = u^\phi/u^t$ of flows moving along Keplerian and free-fall orbits, respectively. The coefficient $\kappa$ is a regulating  parameter. Specifically, when $(\kappa_{\rm ff},\kappa_{\rm K}) = (0,1)$ and $(\kappa_{\rm ff},\kappa_{\rm K}) = (1,0)$, the motion of flows corresponds to pure Keplerian motion and free fall motion, respectively.
    Here, we set to $(\kappa_{\rm ff},\kappa_{\rm K}) = (0.5,0.5)$, which means that the flows are in
    the sub-Keplerian state. Moreover, here the synchrotron emissivity in the disk is modeled by a relativistic thermal (Maxwell-J\"{u}ttner) electron distribution because non-thermal emission is less dominant for the 230 GHz band. In the polarized radiative transfer process, the related coefficients, including emission, absorption, and Faraday effects, are computed using the fitting formulas in \cite{Pandya:2016qfh}. To ensure that the observed flux of thermal synchrotron radiation matches the measured value of Sgr A*, the normalized electron number density $n_{e,0}$ and the electron temperature $T_{e,0}$ are set to $n_{e,0} \approx 10^7 \rm cm^{-3}$ and  $T_{e,0} \approx 10^{11} \rm K$, respectively \cite{Pu_2018}. The detail fiducial values are listed in Table \ref{table:param}.
    \begin{figure}
        \centering
        \includegraphics[width=\textwidth]{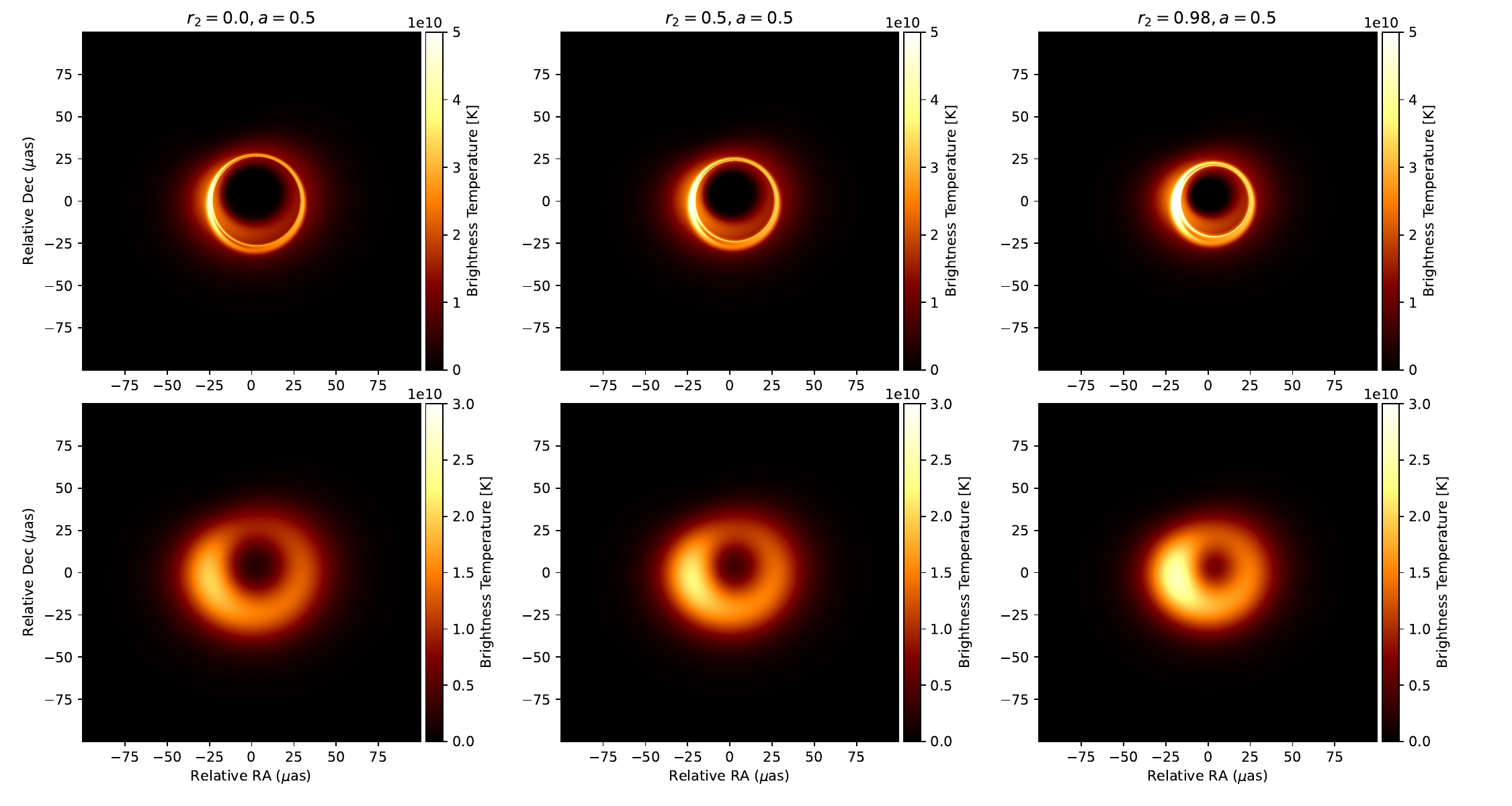}
        \caption{Black hole images (in the top row) and the corresponding blurred images (in the bottom row) at an inclination angle of $30^\circ$ and for a disk thickness $H = 0.3$. The left, middle and right panels are correspond to the cases with $r_2=0.0, 0.5, 0.98$, respectively. Here we set $a=0.5$.}
        \label{fig:1}
    \end{figure}
    In Fig. \ref{fig:1}, we present some black hole images and the corresponding blurred counterparts for Kerr-Sen black holes, based on this semi-analytical optically thin RIAF model.

    \section{EXTRACTED RING FEATURES in images of  Kerr-Sen black holes }
    \label{sec:4}

   To compare theoretical simulations with the photos captured by the EHT under the current observational precision, it is necessary and critical to extract bright ring features from blurred black hole images generated by GRRT. In the REX algorithm employed by the EHT \cite{EventHorizonTelescope:2019ths},
   the ring center $(x_0,y_0)$ is identified by the position that minimizes the normalized radial peak
   dispersion
   \begin{align}
        (x_0,y_0)=\rm argmin\bigg[\frac{\sigma_{\bar{r}(x,y)}}{\bar{r}_{pk}(x,y)}\bigg]_{(x,y)},
    \end{align}
 where $\sigma_{\bar{r}(x,y)}$ is the standard deviation of the values $r_{\rm pk}(\theta; x, y)$.  The quantity $r_{\rm pk}(\theta; x, y)$ is the distance at
which the angular profile assumes its peak brightness and $\bar{r}_{\rm pk}(x,y)$ is the mean
of these peak distances, i.e.,
 \begin{align}
       r_{\rm pk}(\theta; x, y)={\rm argmax}_r[I(r,\theta,x,y)],\quad\quad\quad \bar{r}_{\rm pk}(x,y)= \langle r_{\rm pk}(\theta; x, y)\rangle_{\theta\in [0,2\pi]}.
    \end{align}
$I(r,\theta,x,y)$ is the brightness distribution in the polar coordinates $(r,\theta)$  with the origin at the candidate center of the ring $(x,y)$. The ring diameter $d$ is defined as twice $r_{\rm pk}$ measured from the identified center $(x_0,y_0)$, i.e.,
  \begin{align}
        d = 2r_\text{pk}(x_0,y_0).
    \end{align}
The ring width $w$ corresponds to the azimuthal average of the full width at half maximum (FWHM) of the radial profile,
    \begin{align}
        w=\text{FWHM}[I(r,\theta)-I_{\rm floor}].
    \end{align}
 Here the subtracted value is $I_{\rm floor}=\langle I(r=50\mu as,\theta)\rangle_{\theta}$,  which is introduce to avoid bias in the measurement from the resampled image $I(r,\theta)$ having a nonzero mean floor value outside the ring.
   \begin{figure}
         \centering
        \includegraphics[width=\textwidth]{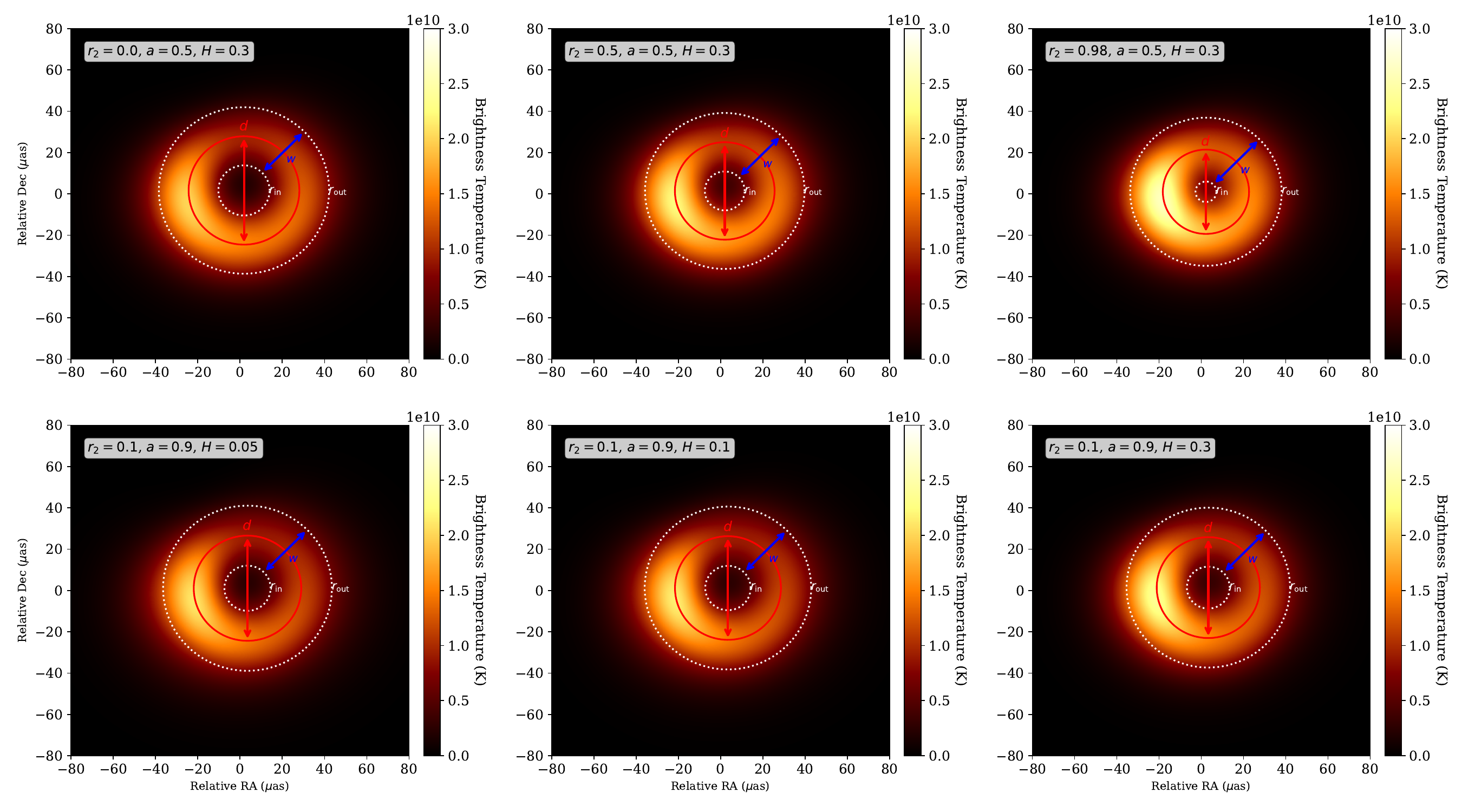}
        \caption{Visualization of extracted ring features of blurred images for different dilaton parameters $r_2$ and disk thickness $H$. Parameter $d$ is the diameter of bright ring; $w$ is the ring width. Parameters $r_{in}$ and $r_{out}$ delimit the radial full width at half maximum (FWHM).}
        \label{fig:2}
    \end{figure}
    \begin{figure}
        \centering
        \includegraphics[width=\textwidth]{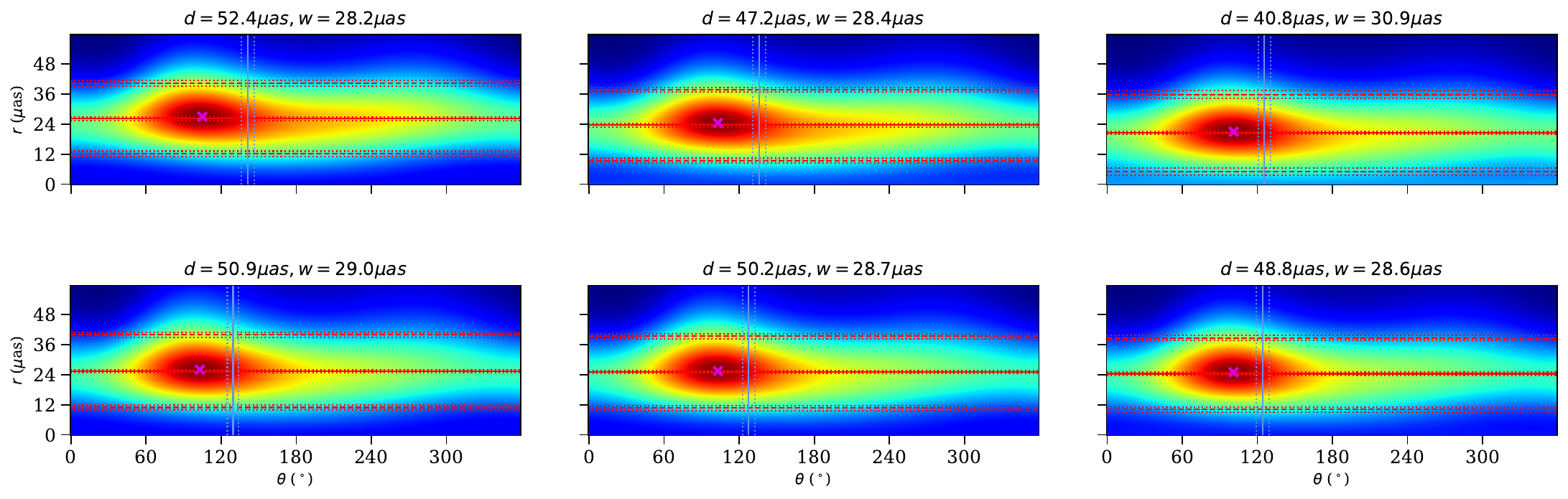}
        \caption{Unwrapped ring profiles of the simulated images in Fig. \ref{fig:2}. The red horizontal lines indicate the measured values of the ring radius $d/2$ and its ring width $(d\pm w)/2$.  The error ranges are marked by dashed lines. Blue vertical solid lines correspond the orientation angle $\eta$ , with their error intervals indicated by blue dashed lines. The purple cross marks the peak brightness in each panels.}
        \label{fig:3}
    \end{figure}
With this REX algorithm, we present the visualization of extracted ring features of blurred images for different dilaton parameters $r_2$ and disk thickness $H$ in Fig. \ref{fig:2}. Moreover, we also present the corresponding unwrapped ring profiles of the simulated images of Kerr-Sen black holes in Fig. \ref{fig:3}.
Our results show that the diameter of the bright ring decreases with the dilaton parameter $r_2$, which is similar to that in the unblurred case shown in Fig. \ref{fig:1}. However, the width and brightness of the ring increase with the dilaton parameter $r_2$.  With the increase of the disk thickness $H$, we find that the diameter and width of the bright ring decrease. The brightness also increases with disk thickness $H$, but the brightness enhancement induced by the disk thickness is less prominent than that driven by the dilaton parameter $r_2$.

From the observation of EHT \cite{sgraI,sgraIV, EHTsgraVI, 2022ApJ...930L..16E,Chael_2018},  the bright ring diameter of the supermassive black hole Sgr A* is $d_{\rm ring}=51.8\pm 2.3~\mu{\rm as}$. Taking into account the current precision of observation, one can compare the bright ring diameter  obtained from the EHT observations with the bright ring diameter  extracted from the simulated blurred images of Kerr-Sen black holes under a Gaussian beam of FWHM $20$ $\mu$.
  \begin{figure}
        \centering
        \includegraphics[width=\textwidth]{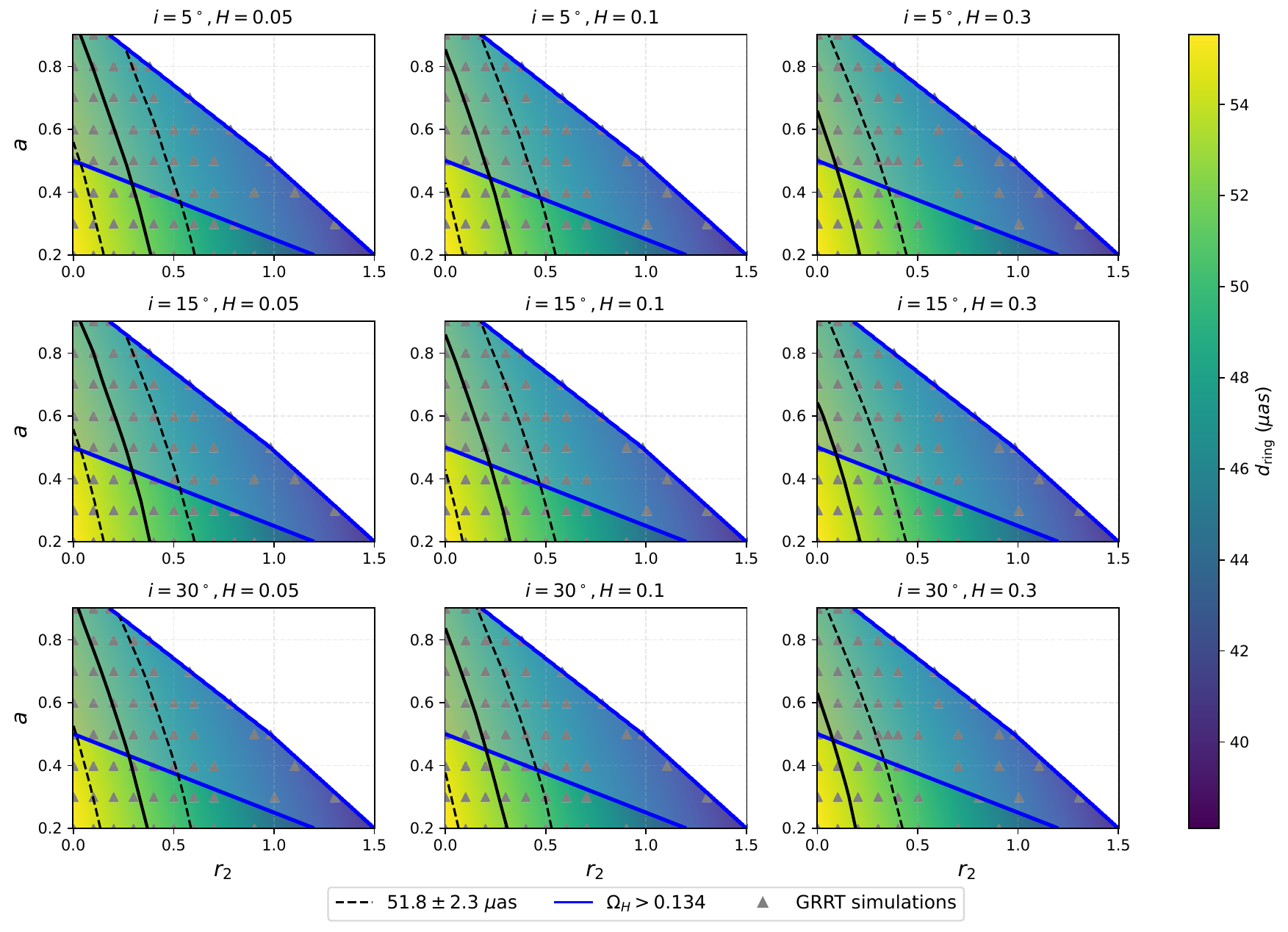}
        \caption{Distribution of the bright ring size on the $r_2-a$ plane for the blurred images of Kerr-Sen black holes with different inclination $i$ and disk thicknesses $H$. The left, middle, and right panels at each row correspond to disk thicknesses of 0.05, 0.1, and 0.3, respectively. The top, middle, and bottom rows correspond to inclination angles of $5^\circ$, $15^\circ$, and $30^\circ$, respectively.}
        \label{fig:4}
    \end{figure}
   In Fig. \ref{fig:4}, we present distribution of the bright ring size on the $r_2-a$ plane for the blurred images of Kerr-Sen black holes with different inclination $i$ and disk thicknesses $H$.
    The gray triangles represent results of the ring diameters extracted by the REx algorithm from the GRRT simulations. The solid and dashed black lines respectively correspond to $d=51.8\mu as $ and its $1\sigma$ uncertainty range (i.e., $\pm 2.3\mu as$), which are from the observation of EHT.
   The blue contours denote regions where  outer event horizon angular momentum satisfies $\Omega_{\rm H}\geq0.134$. The lower limit of $0.134$ corresponds to the outer event horizon angular momentum of Kerr black holes with $a = 0.5$, given that previous analyses of Sgr A* images favor a relatively high black hole spins \cite{2022ApJ...930L..16E}.
    \begin{figure}
        \centering
        \includegraphics[width=\textwidth]{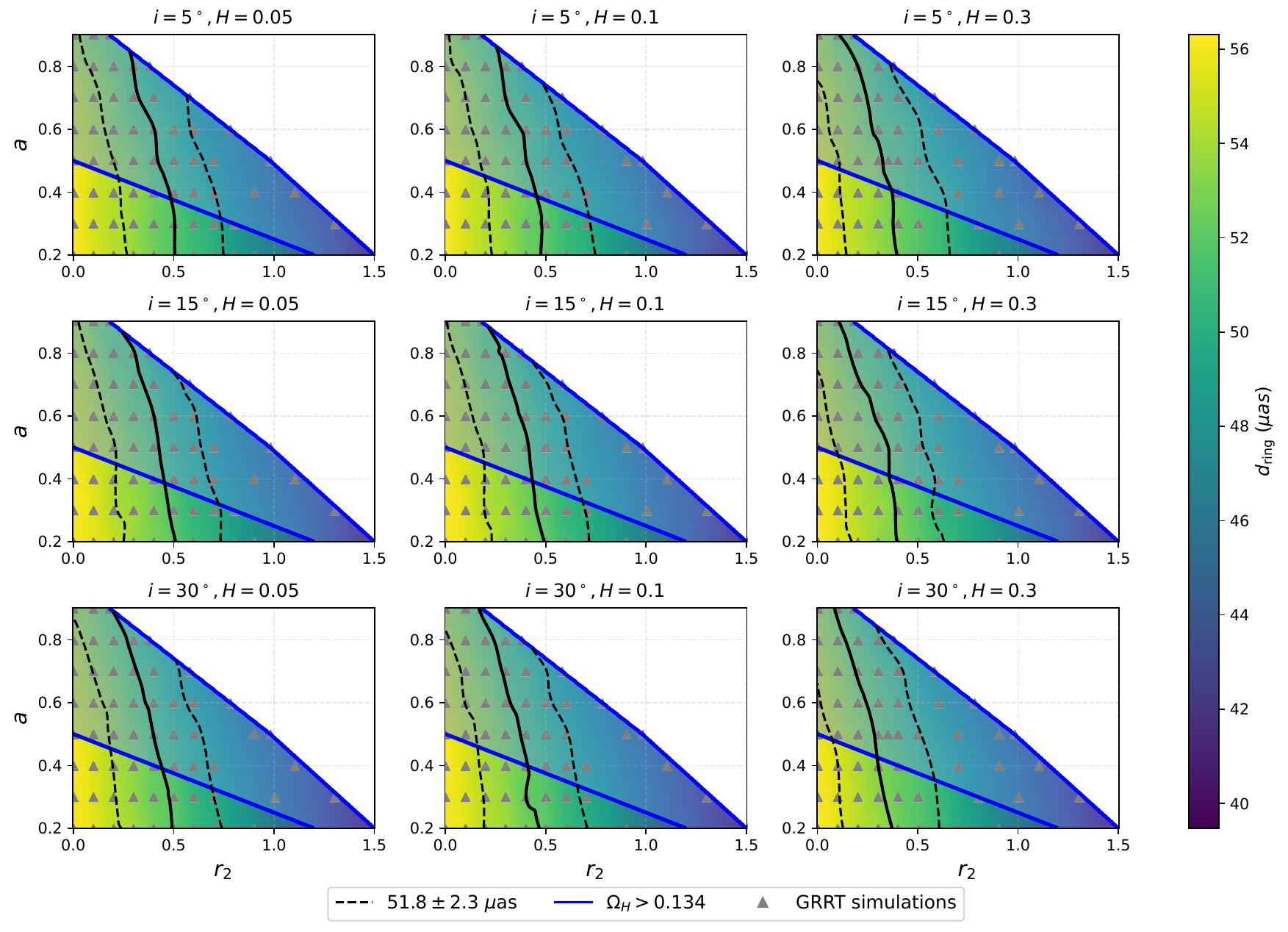}
        \caption{Distribution of the bright ring size on the $r_2-a$ plane for the unblurred images of Kerr-Sen black holes. The parameter settings for the inclination $i$ (rows) and disk thickness $H$ (columns) are the same as in Fig. \ref{fig:4}.}
        \label{fig:5}
    \end{figure}
    Due to the significant uncertainties in the observations by EHT, the allowed parameter space is relatively broad.
    For a fixed disk thickness $H$,  we find that as the inclination angle increases from $5^\circ$ to $30^\circ$, the allowed upper bound of $r_2$ or $a$ remains almost unchanged, while the lower bounds of $r_2$ and $a$ decrease. For a fixed inclination $i$, as $H$ increases, the allowed upper and lower bounds of $r_2$ and $a$ decrease, and the allowed parameter region shifts toward the regime of lower spin and dilaton. This indicates that a larger disk thickness $H$ tends to produce a smaller bright ring diameter. Moreover, we find that effects of the disk thickness $H$ are more significant than those of the observer's inclination angle.
    Fig. \ref{fig:5} corresponds to the unblurred case of Kerr-Sen black hole images. It is shown that the effects of the inclination angle and the disk thickness on the allowed parameter region are similar to those in the blurred case. However, the parameter range obtained from the unblurred case exhibits significant differences from the blurred cases.
    In the unblurred case, both the allowed lower bounds of $a$ and $r_2$ have higher values, the upper bound of $a$ slows down and  the upper bound of $r_2$ increases. These also confirm that the observation resolution significantly affects the precision of constraining black hole parameters.
   \begin{figure}
        \centering
        \includegraphics[width=\textwidth]{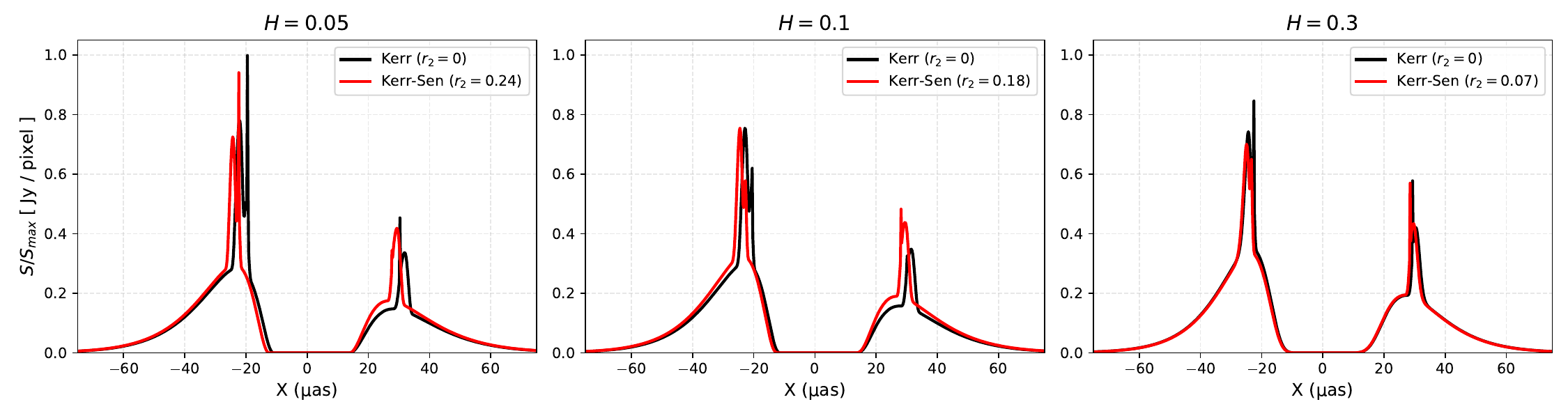}
        \caption{Comparison of the horizontal cross-sectional normalized flux density between images of Kerr black holes and Kerr-Sen black holes with the same ring  diameter $d_{\rm ring}$ for different disk thickness at fixed inclination  $i=30^{\circ}$.}
        \label{fig:7}
    \end{figure}
    \begin{figure}
        \centering
        \includegraphics[width=\textwidth]{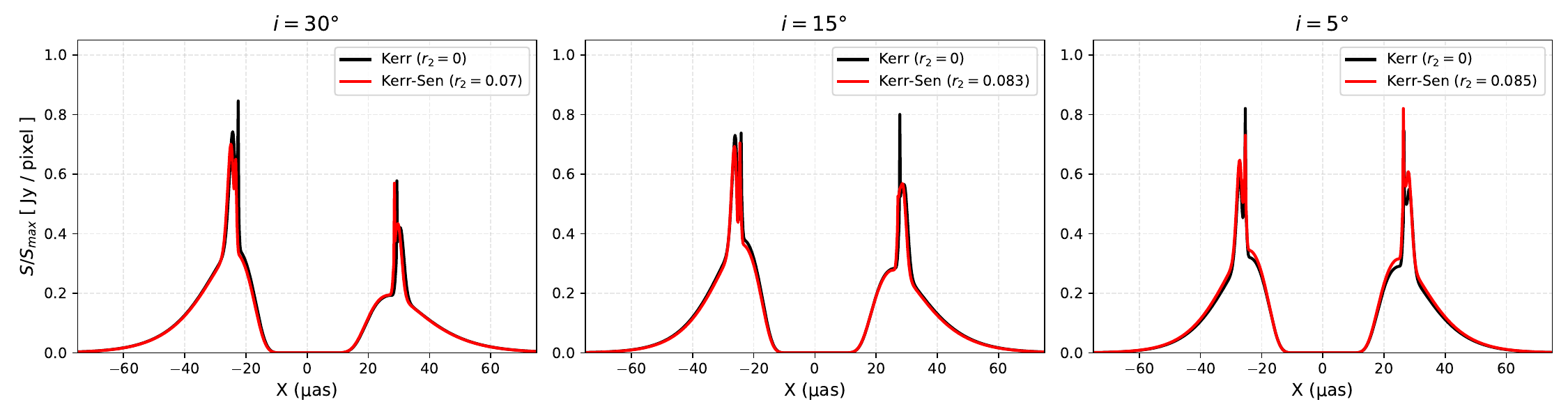}
        \caption{Comparison of the horizontal cross-sectional normalized flux density between images of Kerr black holes and Kerr-Sen black holes with the same ring  diameter $d_{\rm ring}$ for different inclination angle at fixed  disk thickness $H=0.3$.}
        \label{fig:9}
    \end{figure}

   To gain a more detailed understanding of the image of Sgr A* in the context of the Kerr-Sen black hole,  we further explored, in Figs. \ref{fig:7} and \ref{fig:9}, the differences in normalized flux density between the images of Kerr black holes and Kerr-Sen black holes under different thicknesses $H$ and inclinations $i$, with each image having the same ring  diameter $d_{\rm ring} = 51.8~\mu{\rm as}$. Figs. \ref{fig:7} and \ref{fig:9} represents the normalized flux density along the horizontal cross section of the images. As the disk thickness increases, the difference in profiles between the images of Kerr black holes and Kerr-Sen black holes decreases, and the profiles almost coincide when  $H=0.3$. This means that for black hole images with the same ring diameter $d_{\rm ring}$,it is more difficult to distinguish between Kerr black hole and Kerr-Sen black hole based on the brightness distribution when the accretion disk has a larger thickness. In addition, from Fig. \ref{fig:9}, we find that the difference in profiles decreases with the inclination angle $i$,  such that  the profiles almost coincide when  $i=30^{\circ}$. Comparing with Figs. \ref{fig:7} and \ref{fig:9}, we can find that effect of the accretion disk thickness on the difference in profiles is slightly more pronounced than that of the inclination angle.  Therefore, to observe these differences and further resolve the degeneracy in images, higher-resolution observations are required, such as those that will be obtainable from future ngEHT observations \cite{tiede2022measuringphotonringsngeht,Johnson_2023}.

    \section{Polarization structures in images of Kerr-Sen black holes }
    \label{sec:5}

\begin{figure}
        \centering
        \includegraphics[width=\textwidth]{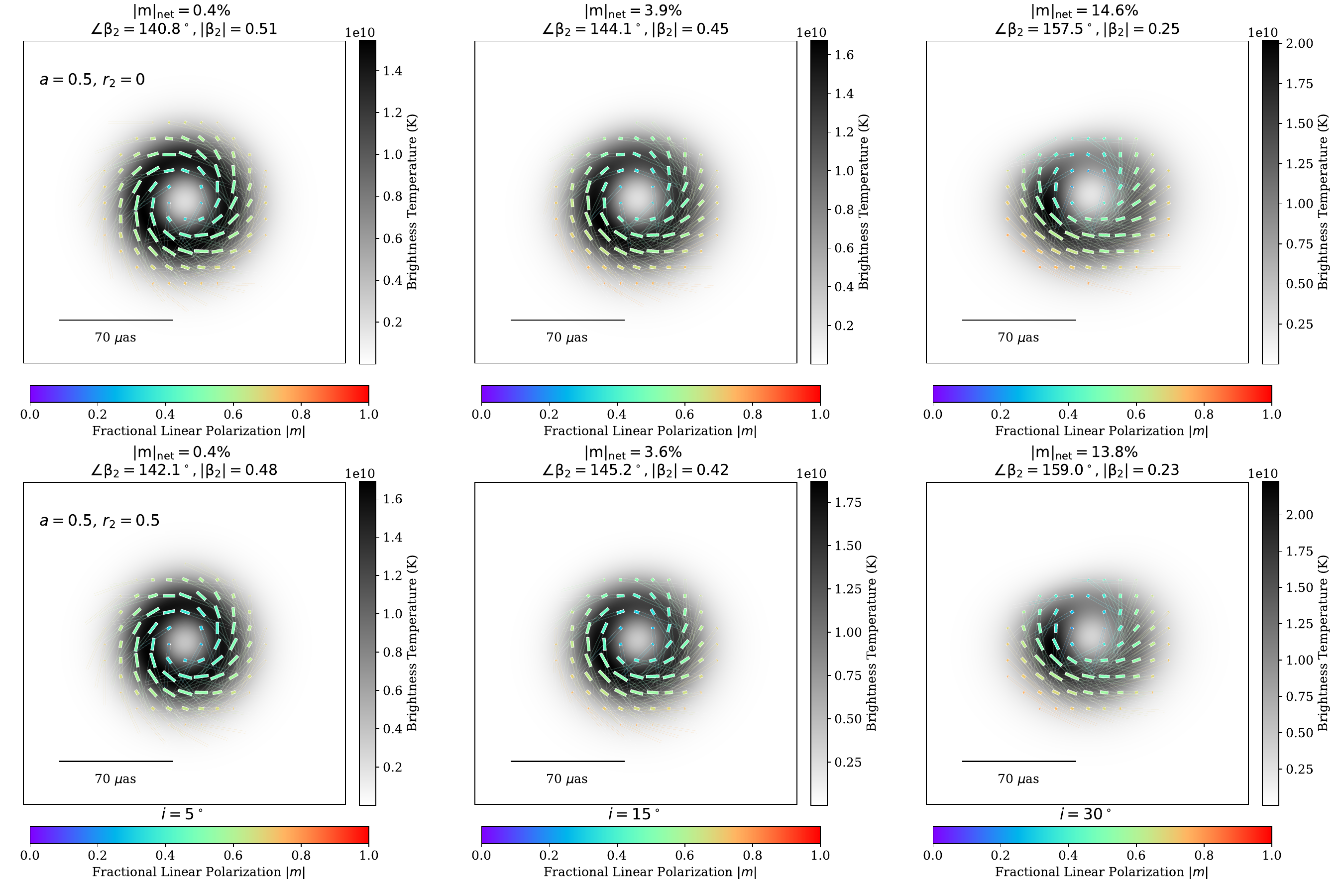}
        \caption{Polarization images of Kerr-Sen black holes with different inclination angles for fixed disk thickness $H=0.3$ and $a=0.5$. The upper and bottom rows correspond to the cases with the dilaton parameter $r_2=0$ and $r_2=0.5$, respectively. The grayscale represents the total intensity, while the short lines indicate the degree and direction of linear polarization. The color of the lines represents the magnitude  of partial linear polarization.}
        \label{fig:10}
    \end{figure}

   In general, the polarization information carried by light beams can be described by the Stokes parameters $I$, $Q$, $U$, and $V$ \cite{1996AAS..117..161H,Smirnov_2011}. The parameter $I$ represents the total emission, and other parameters $Q$, $U$, and $V$ describe the polarized portion of the emission. Using these quantities, the electric-vector position angle (EVPA) is defined as
    \begin{equation}
        {\rm EVPA} = \frac{1}{2}{\rm arctan}(\frac{U}{Q}),
    \end{equation}
    and the linear polarization fractions $\rm |m|_{net}$ can be expressed as
    \begin{equation}         |m|_{\mathrm{net}}=\frac{\sqrt{\left(\sum_jQ_j\right)^2+\left(\sum_jU_j\right)^2}}{\sum_jI_j},
    \end{equation}
    where the sums are taken over all pixels $j$ in the images.
    In Fig. \ref{fig:10}, we employ a simplified vertical magnetized Sub-Keplerian flow model to simulate the polarization images of Kerr-Sen black holes with different inclination angles for fixed disk thickness $H=0.3$.
    From Fig. \ref{fig:10}, we can find that there is a similar twisting polarization pattern for different values of the dilaton parameter $r_2$.

    To further quantitatively characterize the polarization mode distribution, one can define a set of quantitative indices $\beta_m$ for the EVPA patterns through the scalar product of the complex polarization quantity $P=Q+iU$  and a set of basis functions \cite{ehtVII,m87VIII,m87VIII,Palumbo_2020,Chael_2023}, i.e.,
    \begin{equation}
    \beta_m=\frac{1}{I_{\rm ann}}\int_{r_{\rm min}}^{r_{\rm max}}\int_0^{2\pi}P(r,\phi)e^{im\phi}r\mathrm{~d}\phi\mathrm{~d}r,
    \end{equation}
    where the chosen basis functions $P_b = e^{im\phi}$ represent the possible spatial modes of EVPA variation with the azimuthal angle $\phi$ in the image plane, and $m$ is the corresponding mode order \cite{m87VIII,Palumbo_2020,Chael_2023}. The quantity $I_{\rm {ann}}$ has a form
    \begin{equation}
    I_{\rm {ann}}=\int_{r_{\rm min}}^{r_{\rm max}}\int_{0}^{2\pi}I(r,    \phi)r\mathrm{d}\phi\mathrm{d}r.
    \end{equation}
   where the inner radius $r_{\rm min}$ and outer radius $r_{\rm max}$ correspond to the inner and outer edges of the emission ring shown in the image, respectively.
   In general, the $\beta_m$ coefficients are complex numbers. The magnitude $|\beta_m|$ quantifies the degree of $m$-fold rotational symmetry in the EVPA distribution around the ring, while the argument $\angle\beta_m$ measures the average orientation of the symmetric component, relative to a reference pattern with vertical EVPA at image north \cite{m87VIII,Palumbo_2020}. The model $m = 2$ serves as a robust discriminator for distinguishing between MAD and SANE flows in terms of black hole images.
    \begin{figure}
        \centering
        \includegraphics[width=\textwidth]{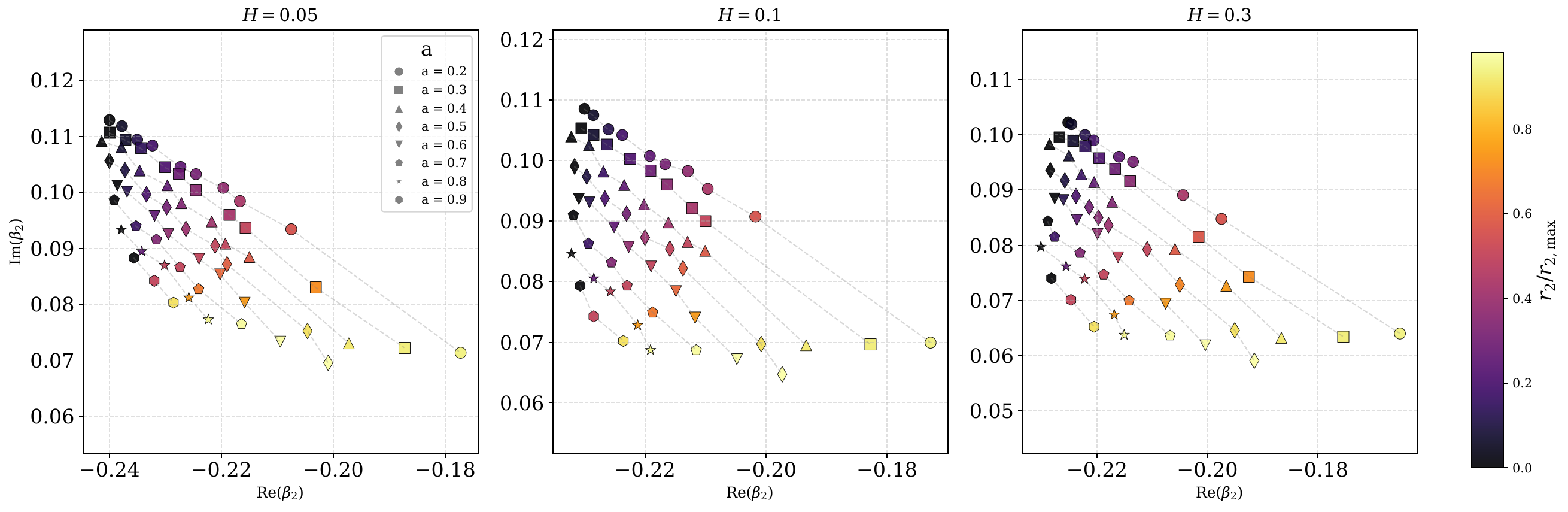}
        \caption{Variation of the complex $\beta_2$ coefficient with the normalized dilaton parameter $r_2 / r_{2\text{max}}$ for fixed inclination $i=30^{\circ}$ and different disk thickness $H$.}
        \label{fig:11}
    \end{figure}
    \begin{figure}
        \centering
        \includegraphics[width=\textwidth]{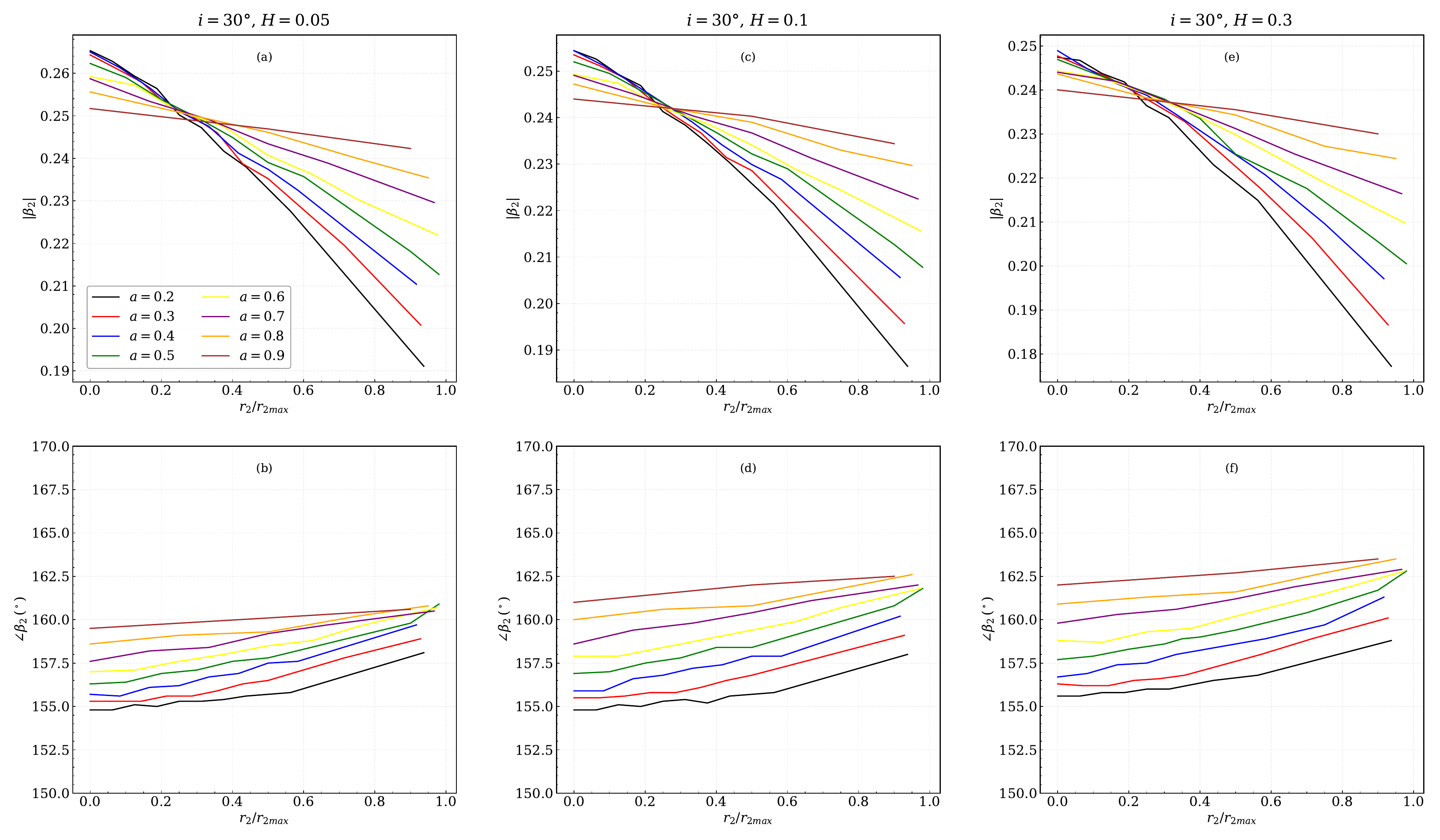}
        \caption{Changes of the magnitude  and phase of the complex coefficient $\beta_2$ with the normalized dilaton parameter $r_2 / r_{2\text{max}}$ for fixed inclination $i=30^{\circ}$ and different disk thickness $H$.}
        \label{fig:12}
    \end{figure}
   Figs. \ref{fig:11} and \ref{fig:12} present changes of the complex mode $\beta_2$ with the normalized dilaton parameter $r_2 / r_{2\text{max}}$ for fixed inclination $i=30^{\circ}$ and different disk thickness $H$ after applying a 20 $\mu as$ Gaussian blur to the original simulated images of SgrA*.
   As the dilaton parameter increases, the real part $Re \beta_2$ increases and the imaginary part $Im \beta_2$ decreases. Moreover, it is found that the real part $Re \beta_2$  is always negative, which means the phase angle is in the range $\angle\beta_2\in (\pi/2,\pi)$. The magnitude  $|\beta_2|$ generally exhibits a declining trend as $r_2 / r_{2\text{max}}$ increases.  Particularly in the case of a low spin parameter,  the decline is more pronounced. In contrast, in the case with large spin parameter, the decline is slower or even tends to flatten.  The angle $\angle\beta_2$ increases monotonically with $r_2 / r_{2\text{max}}$, while its rate of increase decreases with the black hole spin.
   As the disk thickness $H$ increases, the magnitude  $|\beta_2|$ decreases slightly, and the phase angle $\angle\beta_2$ increases slightly. However, these effects are much weaker than those induced by the dilaton parameter. These results are beneficial in understanding black hole images with accretion disks and the gravity of Einstein-Maxell-Dilaton-Axion theory.

    \section{Summary}
    \label{sec:6}

    We have investigated the bright ring features and polarization structures in the images illuminated by the 230 GHz thermal synchrotron emission from the RIAF around Kerr-Sen black holes. Our results reveal that an increase in the dilaton parameter leads to a shrinking of the bright ring, accompanied by enhancements in both its width and brightness.  Actually, these effects can be explained by the fact that the dilaton parameter $r_2$ weakens the black hole's gravity because the outer horizon radius $r_+$ decreases with the dilaton parameter $r_2$ due to $\frac{dr_{+}}{dr_2}=-\frac{r_+}{2\sqrt{(M-\frac{r_2}{2})^2-a^2}}<0$.  Therefore, compared to black holes without dilaton parameter $r_2$, only photons passing closer to the black hole can fall into it, thereby reducing the diameter of the bright ring. Simultaneously, it allows light emitted from broader regions on the accretion disk to reach the observer, leading to an increase in both the brightness and width of the bright ring. In parallel, we find that as the disk thickness grows, both the diameter and width of the bright ring decrease. While the brightness also increases with the disk thickness $H$, this enhancement is less prominent than that driven by the dilaton parameter $r_2$.  We also present the allowed ranges of black hole parameters through comparing with the EHT observational data. For blurred images convolved with a Gaussian kernel $20\,\mu\mathrm{as}$, we find that for a fixed disk thickness $H$, increasing the inclination angle $i$ reduces the lower bounds of parameters $r_2$ and $a$, while their upper bounds remain largely unchanged. For a fixed inclination $i$,  increasing  the disk thickness $H$ leads to decreasing the lower bounds of $r_2$ and $a$, and shifting the overall allowed parameter space toward the region with the lower values of $a$ and $r_2$. These imply that the disk thickness $H$ exerts a stronger influence on the parameter space than the observer's inclination $i$, and a larger disk thickness $H$ tends to yield a smaller bright ring diameter. In contrast, for the unblurred images, we find that there exhibits the higher bottom bounds for $a$ and $r_2$,  while the upper bound of $a$ slows down and the upper bound of $r_2$ increases. These also confirm that the observation resolution affects significantly the precision of constraining black hole parameters.

    We also analyzed the influence of the dilaton parameter $r_2$ on polarization structures in Kerr-Sen black hole images by quantifying EVPA rotational symmetry using the method of $\beta_2$. Our results show that as $r_2$ increases, the real part $Re\beta_2$ increases and the imaginary part $Im\beta_2$ decreases, but the magnitude  $|\beta_2|$ generally exhibits a declining trend. The angle phase $\angle\beta_2$ increases monotonically and its increasing rate decreases with the black hole spin. In refs.\cite{Wong:2025zuh,10.1093/mnras/staf200}, one can find that the phase angle $\angle\beta_2$ from the 230 GHz images from the radiative simulations can be approximately fitted with a simple function $\angle\beta_2=-2\arctan \frac{C_0}{|\Omega_H|}+C_1$, where $\Omega_H=\frac{a}{2r_+}$ is the angular velocity at outer event horizon, $C_0$ and $C_1$ are  numerical parameters. Consequently, one has the value  $|\beta_2|\sim \sqrt{1+\tan^2\angle\beta_2}\sim \frac{2\Omega_H}{\Omega^2_H-C^2_0}$. With these fitted formulas, one can find that an increase in angular velocity triggered by the dilaton parameter results in  a corresponding increase in the angle $\angle\beta_2$ and a decrease in the magnitude  $|\beta_2|$. This further highlights the importance of the frame-dragging effect with respect to the polarization structure in black hole images. The disk thickness $H$ results in that both the magnitude  $|\beta_2|$ decreases slightly, and the phase angle $\angle\beta_2$ increases slightly, but its effects are much weaker than those from the dilaton parameter.

    \vspace{2cm}

    \centering{\bf Acknowledgments}

    This work was supported by the National Natural Science Foundation of China under Grant No.12275078, 11875026, 12035005, 2020YFC2201400, and the innovative research group of Hunan Province under Grant No. 2024JJ1006.

    \bibliography{kerrmog}

\end{document}